\documentclass[10pt,conference,a4paper]{IEEEtran}
\usepackage{times}

\usepackage[english]{babel}
\usepackage[T1]{fontenc}
\usepackage[utf8]{inputenc}

\usepackage{amsmath,amssymb}
\usepackage{xcolor}
\usepackage{colortbl}
\usepackage{booktabs}
\usepackage{graphicx}
\usepackage{tabularx}

\usepackage{pgf}
\usepackage{pgffor}
\usepgflibrary{plothandlers}

\graphicspath{{images/}}

\title{Reducing Randomness of Non-Regular Sampling Masks for Image Reconstruction}
\author{%
{Markus Jonscher, Jürgen Seiler, Thomas Richter, André Kaup}
{}
\vspace{1.6mm}\\
\fontsize{10}{10}\selectfont\itshape
Multimedia Communications and Signal Processing \\
Friedrich-Alexander-Universität Erlangen-Nürnberg, Cauerstr. 7, 91058 Erlangen, Germany \\
\fontsize{9}{9}\selectfont\ttfamily\upshape
\{jonscher,seiler,richter,kaup\}@LNT.de

\vspace{1.2mm}\\
\fontsize{10}{10}\selectfont\rmfamily\itshape

\fontsize{9}{9}\selectfont\ttfamily\upshape
}
\begin{document}
\maketitle

\begin{figure}[b]
\parbox{\hsize}{\em

978-1-4799-6139-9/14/\$31.00 \ \copyright 2014 IEEE
}\end{figure}

\begin{abstract}
Increasing spatial image resolution is an often required, yet challenging task in image acquisition. 
Recently, it has been shown that it is possible to obtain a high resolution image by covering a low resolution sensor with a non-regular sampling mask. Due to the masking, however, some pixel information in the resulting high resolution image is not available and has to be reconstructed by an efficient image reconstruction algorithm in order to get a fully reconstructed high resolution image. 
In this paper, the influence of different sampling masks with a reduced randomness of the non-regularity on the image reconstruction process is evaluated. Simulation results show that it is sufficient to use sampling masks that are non-regular only on a smaller scale. These sampling masks lead to a visually noticeable gain in PSNR compared to arbitrary chosen sampling masks which are non-regular over the whole image sensor size. At the same time, they simplify the manufacturing process and allow for efficient storage.
\\[1\baselineskip]
\end{abstract}



\section{Introduction}
\label{sec:introduction}
In most imaging applications, image sensors with high resolution (HR) are required. By using existing low resolution (LR) imaging systems and applying super resolution (SR) techniques \cite{Park2003}, an HR image can be obtained without replacing existing LR sensors by models with higher resolution. SR is typically based on registration, interpolation, and deblurring and usually, multiple LR images are necessary in order to obtain an HR image. There are also SR techniques that may be applied to single LR images \cite{Zhang2012}. 

Recently, a different approach for increasing the camera resolution has been proposed in \cite{Schoeberl2011a}. The idea is to cover an LR sensor with a non-regular sampling mask and reconstruct an HR image afterwards.
\begin{figure}[t]
	\centering
	\def\svgwidth{\columnwidth}	
	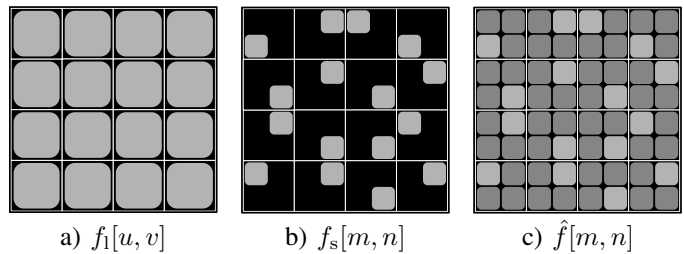	
	\caption{An LR sensor (a) is covered with a non-regular sampling mask (b). Employing FSE for reconstructing missing pixels (dark gray), an image with higher resolution can be obtained (c).}
	\label{fig:reconstruction_pipeline}
\end{figure}
The principle of this idea is illustrated in Fig.~\ref{fig:reconstruction_pipeline}. An LR sensor with regular large pixels (Fig.~\ref{fig:reconstruction_pipeline}a) is covered with a sampling mask so that a non-regular sampling pattern on an HR grid occurs (Fig.~\ref{fig:reconstruction_pipeline}b). The light gray color denotes the area sensitive to light.
Each large pixel of the LR sensor is divided into four quadrants where three of them are randomly covered and as a consequence not sensitive to light.
Since due to the masking only $25 \%$ of the pixel size is available, this is called $1/4$~sampling. Although all directly sampled areas are randomly distributed on the HR grid, the underlying image sensor architecture with all the electronic parts is still regular. Missing pixels on the HR grid have to be reconstructed by a suitable reconstruction method in order to obtain a fully reconstructed HR image (Fig.~\ref{fig:reconstruction_pipeline}c). As suggested in \cite{Schoeberl2011a}, a high image quality can be achieved by using Frequency Selective Extrapolation (FSE) \cite{Seiler2010}.
\begin{figure*}[!ht]
	\centering
	\def\svgwidth{\textwidth}	
	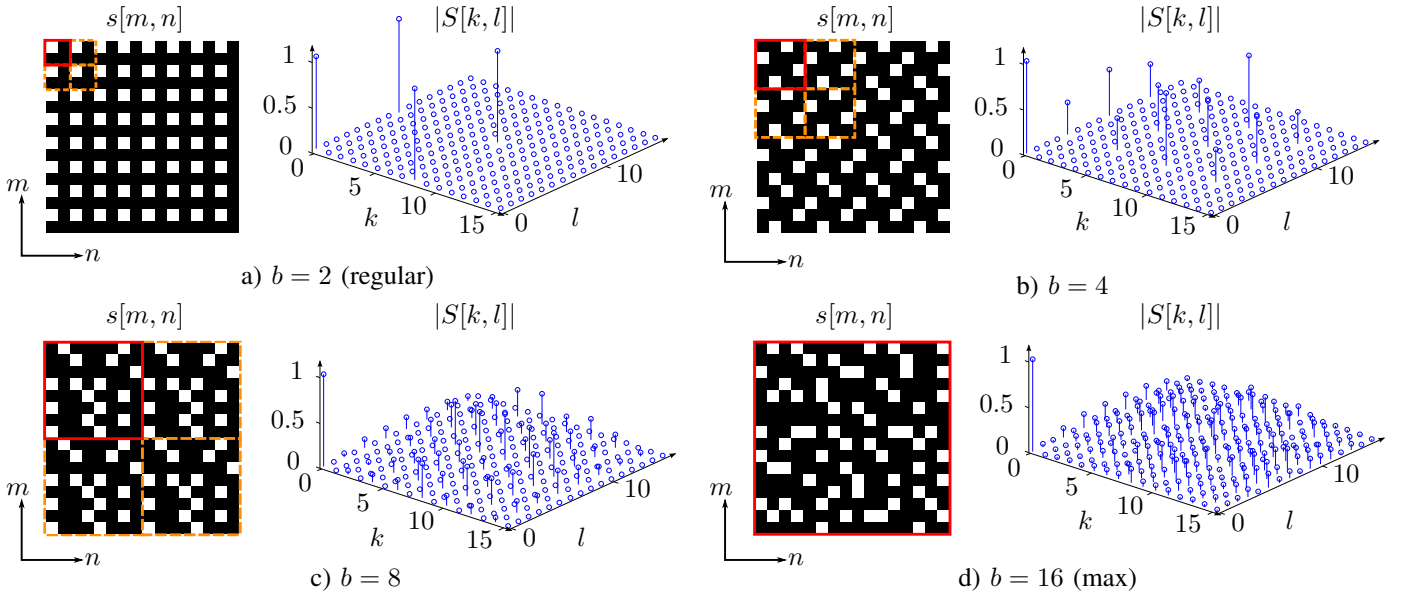
	\caption{Different sampling masks $s[m,n]$ and their corresponding normalized amplitude spectra $\left|S[k,l]\right|$. White areas in $s[m,n]$ are sensitive to light. Blocks of size $b$ are marked by a red box with solid lines and their first adjacent repetitions by orange boxes with dashed lines.}
	\label{fig:sampling_masks}
	\vspace*{-0.5cm}
\end{figure*}

It has been shown in \cite{Hennenfent2007} and \cite{Maeda2009} that sampling image data on a non-regular grid causes less aliasing. This effect is visible from evolution, since rod/cone cells in human eyes are spatially randomly distributed \cite{Roorda1999}. Thus, using the image acquisition system mentioned above and due to the non-regular subsampling of the area sensitive to light, aliasing in the reconstructed image can be reduced.
In this paper, the effects of non-regular subsampling on image reconstruction and the influence of the non-regularity of the sampling masks are evaluated. Thereby, sampling masks are generated where the non-regularity only occurs on smaller blocks which are then repeated to match the image sensor size. This may benefit the manufacturing process of such a sampling mask, since only a small template has to be stored. Throughout this work, non-regular always means that the samples are distributed non-regularly over the whole sensor size employing the idea of $1/4$~sampling.

The paper is organized as follows: The next section shows different sampling architectures. 
Extensive simulations are presented in Section~\ref{sec:simulations} and the results are given in Section~\ref{sec:analysis}.
Section~\ref{sec:conclusion} concludes the paper.


\section{Sampling Architectures}
\label{sec:sampling}
Covering an image sensor with a physical mask leads to a subsampled image regarding the HR grid. All sampling masks used in this paper employ $1/4$~sampling. This corresponds to a subsampling of factor 2 in both spatial dimensions which are depicted by $(m,n)$.
The image $f_{\mathrm{l}}[u,v]$ that a low resolution sensor would acquire (Fig.~\ref{fig:reconstruction_pipeline}a) is regarded on the LR grid, depicted by the spatial coordinates $(u,v)$.
Masking can be expressed as a multiplication of the signal $f[m,n]$ that an HR sensor would acquire with a binary sampling mask $s[m,n]$. This results in a captured image
\begin{equation}
	f_{\mathrm{s}}[m,n] = s[m,n] \cdot f[m,n]
	\label{eq:masking}	
\end{equation}
with all directly sampled pixels non-regularly distributed on a grid with twice the resolution in both dimensions compared to an LR sensor. These pixels hold exactly those values that an HR sensor would acquire at the corresponding positions. There are several possibilities to design a sampling mask $s[m,n]$. For all generated masks, $1/4$~sampling is applied on smaller blocks of size $b$ which are then repeated to match the image sensor size. These blocks are called templates.
This way, completely regular masks (Fig.~\ref{fig:sampling_masks}a) where $b = 2$ are possible and also masks that are non-regular over the whole sensor size (Fig.~\ref{fig:sampling_masks}d) where in this example $b = 16$. $b=\max$ has been utilized in \cite{Schoeberl2011a}. Examples for other sampling masks with blocks of size ${b=4}$ and ${b=8}$ are shown in Fig.~\ref{fig:sampling_masks}b,c, where the template is marked by a red box with solid lines and the first adjacent repetitions are marked by orange boxes with dashed lines. 

Spatial sampling in (\ref{eq:masking}) corresponds to a circular convolution in frequency domain and can be expressed as
\begin{equation}
	\mathcal{F}\left\lbrace f_{\mathrm{s}}[m,n] \right\rbrace = S[k,l] \circledast \mathcal{F}\left\lbrace f[m,n] \right\rbrace
\end{equation}
with the two-dimensional discrete Fourier transform $\mathcal{F}\left\lbrace\cdot\right\rbrace$. Regarding the normalized amplitude spectra \mbox{$S[k,l] = \mathcal{F}\left\lbrace s[m,n] \right\rbrace$} of different sampling masks in Fig.~\ref{fig:sampling_masks}, it can be seen that in case of regular subsampling four equally dominant peaks occur. Thus, the subsampled signal has spectral overlapping which leads to the well-known phenomenon of aliasing. For non-regular $1/4$~sampling over the whole sensor size, as shown in Fig.~\ref{fig:sampling_masks}d, the signal has only one dominant peak and all other frequencies contribute as weak noise. Masks with smaller blocks of size $b$ also have only one dominant peak. Other frequencies, however, may contribute more than in the completely non-regular case, but always less than in the regular case. For the idea of $1/4$~sampling and the reconstruction of incompletely masked images it is important to design a sampling mask that guarantees the fewest aliasing and supports the reconstruction.

\section{Measurements}
\label{sec:simulations}
In this section, different sampling architectures mentioned in Section \ref{sec:sampling} are evaluated regarding their influence on image reconstruction. Therefore, two different sets of test images are used. The first set consists of $24$ images of size $768 \times 512$ pixels from the KODAK image library. These images are widely used for comparing image processing techniques. The second set consists of $100$ images of size $1200 \times 1200$ pixels, where only the first $24$ images are used. This archive is intended for scientific purposes and was published by TECNICK \cite{Asuni2011}. The images of both test sets are converted to grayscale.
Sampling masks are generated from regular (Fig.~\ref{fig:sampling_masks}a, $b=2$) to non-regular over the whole image sensor size (Fig.~\ref{fig:sampling_masks}d, $b=\max$). Up to now, the latter has been used in \cite{Schoeberl2011a} just to demonstrate the ability of $1/4$~sampling to reconstruct HR images from LR image sensors. Furthermore, sampling masks with blocks of size $b = 2^i$ and $i=2,\ldots,7$ are generated as presented in Section \ref{sec:sampling}.
For $1/4$~sampling, there are 
\begin{equation}	
	N_{\mathrm{masks}} = 4^{\frac{b^2}{4}}
\end{equation}
possible solutions to design a sampling mask for a specific block of size $b$.
$b = 2$ would result in four regular masks. For $b = 4$, there are $256$ possible ways of generating a sampling mask including those masks that are regular ($b=2$) or have regular "super-pixels" (see Fig.~\ref{fig:different_masks}d). The latter may occur when four resulting sampling points of $2 \times 2$ regular large pixels form one $2 \times 2$ "super-pixel" on the HR grid. For $b = 8$, there are already around $4.3$ billion possibilities which is, especially for even higher values of $b$, computational not feasible. Therefore, for each block of size $b$, $256$ randomly picked sampling masks are selected for simulation. In the end, seven sets consisting of $256$ different sampling masks and one set consisting of four regular masks ($b=2$) are used. 
As suggested in \cite{Schoeberl2011a}, Frequency Selective Extrapolation (FSE) \cite{Seiler2010} may be applied to obtain a reasonable HR image quality. Additionally, linear interpolation (LIN) and steering kernel regression (SKR) \cite{Takeda2007} are utilized to compare the image reconstruction quality and the influence of the sampling masks on different reconstruction methods.
Each of the test images is multiplied with one of the corresponding $256$ sampling masks. It is then reconstructed and an average PSNR over all $24$ images for each set is computed. 

\section{Results and Discussion}
\label{sec:analysis}
\begin{figure}[]
	\vspace*{-0.45cm}
	\centering
	\def\svgwidth{\columnwidth}	
	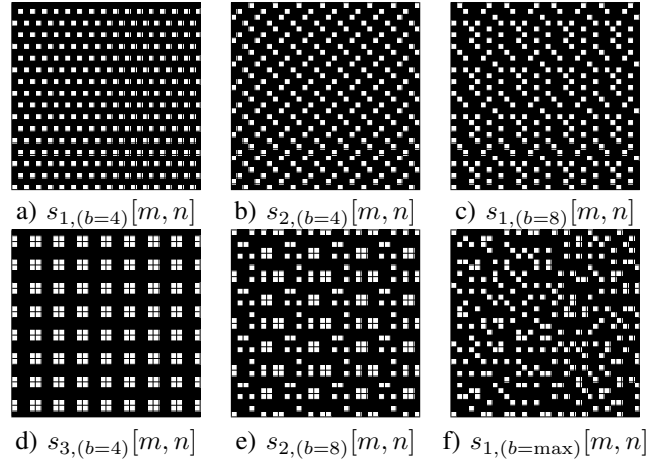
	\vspace*{-0.8cm}
	\caption{Various non-regular sampling masks that have been utilized.}
	\label{fig:different_masks}
	\vspace*{-0.3cm}
\end{figure}

Table \ref{tab:psnr_results} shows PSNR results in dB for several non-regular sampling masks with different blocks of size $b$ and a comparison between LIN, SKR, and FSE for the two test data sets. All values are measured in dB and taken for the particular mask that gives the best results for each image reconstruction method.
LIN performs best for a sampling mask with $b=4$ which is shown in Fig.~\ref{fig:different_masks}a. Although this sampling mask is regular but not separable, an average gain of $0.62$~dB is possible compared to a completely regular sampling mask which is separable (Fig.~\ref{fig:sampling_masks}a).
In Fig.~\ref{fig:different_masks}b, the sampling mask is illustrated that performs well for SKR. It also has blocks of size $b=4$ but it appears more random. Compared to a regular subsampling, this leads to a gain of $0.57$~dB.
Using a non-regular sampling mask for FSE gives a gain of $0.88$~dB compared to a regular sampling mask. It can also be seen from the table that FSE outperforms LIN and SKR by up to $0.97$~dB concerning the overall reconstruction quality. This can be achieved by using a sampling mask with $b=8$. The best mask for this block size is displayed in Fig.~\ref{fig:different_masks}c. A sampling mask which performs worst for all regarded reconstruction methods is shown in Fig.~\ref{fig:different_masks}d. Here, the before mentioned "super-pixel" can be seen. In Fig.~\ref{fig:different_masks}e, another unfavorable mask is shown. Here, clumps and constellations like "super-pixel" occur which are not suited for the reconstruction process. 

Similar results have been developed in \cite{LI2008} but rather for completely random sampling patterns and not for $1/4$~sampling. It has been stated that randomized sampling patterns have to be optimized in a way that an uniformity in the distribution of the samples is guaranteed. When employing $1/4$~sampling, however, it just has to be ensured that clumps and other constellations in sampling masks are avoided, since these masks per se are more uniform than totally random generated masks. Comparing the mask that gives the best results (Fig.~\ref{fig:different_masks}c) to $1/4$~sampling mask which are non-regular over the whole sensor size (Fig.~\ref{fig:different_masks}f), there is little difference in the structure of the sampling mask. The difference between the best and the worst non-regular sampling mask with $b=\max$ is only $0.03$~dB which means that the probability is very high to choose such a sampling mask. Regular masks are also included in $b=\max$ but it is very unlikely to choose them.
\renewcommand{\arraystretch}{1.2}
\begin{table}[t]
	\caption{PSNR results in dB for non-regular sampling masks with different blocks of size $b$ over two test sets.}
	\label{tab:psnr_results}
	\vspace*{-0.2cm}
	\centering
	\begin{tabularx}{\columnwidth}{X|ccc|ccc}
		\toprule
		         &               \multicolumn{3}{c|}{KODAK}               &                \multicolumn{3}{c}{TECNICK}               \\
		         &       LIN        &       SKR        &       FSE        &         LIN        &       SKR        &       FSE        \\ \midrule
		$b=2$    &     $27.96$      &     $27.84$      &     $28.11$      &       $30.64$      &     $30.05$      &     $31.07$      \\
		$b=4$    & $\mathbf{28.69}$ & $\mathbf{28.14}$ &     $28.99$      &   $\mathbf{31.15}$ & $\mathbf{30.88}$ &     $31.76$      \\
		$b=8$    &     $27.88$      &     $27.92$      & \cellcolor{blue!15}{$\mathbf{29.11}$} &       $30.40$    &     $30.68$      & \cellcolor{blue!15}{$\mathbf{31.82}$} \\
		$b=16$   &     $27.54$      &     $27.73$      &     $28.96$      &       $30.07$      &     $30.51$      &     $31.69$      \\
		$b=32$   &     $27.40$      &     $27.64$      &     $28.87$      &       $29.92$      &     $30.39$      &     $31.59$      \\
		$b=64$   &     $27.35$      &     $27.58$      &     $28.83$      &       $29.86$      &     $30.33$      &     $31.53$      \\
		$b=128$  &     $27.32$      &     $27.57$      &     $28.82$      &       $29.83$      &     $30.32$      &     $31.52$      \\
		$b=\max$ &     $27.31$      &     $27.55$      &     $28.80$      &       $29.81$      &     $30.30$      &     $31.50$      \\ \bottomrule
	\end{tabularx}
\end{table}
\begin{figure*}[ht]
	\centering
	\def\svgwidth{\textwidth}
	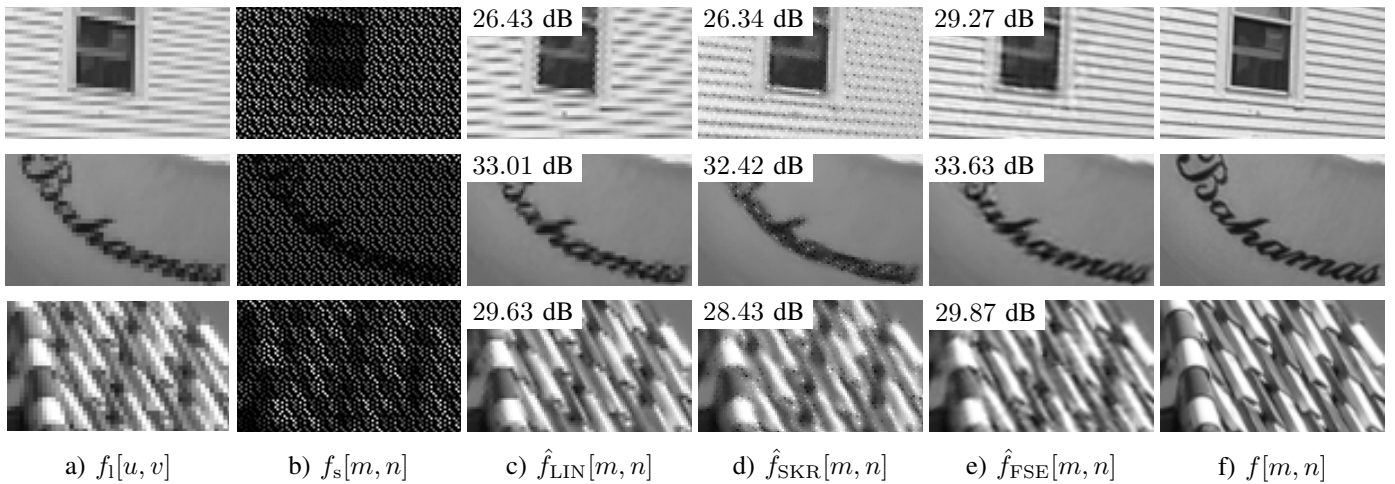
	\vspace*{-0.6cm}
	\caption{Image detail examples from the KODAK and TECNICK image library. From top to bottom: {\it{house}}, {\it{text}}, {\it{roof}}; all reconstructed for the best sampling mask for each reconstruction method. $f_s[m,n]$ shows exemplarily the mask used for FSE.}
	\label{fig:image_examples}
	\vspace*{-0.1cm}
\end{figure*}
\begin{figure*}[t]
	\centering
	\def\svgwidth{\textwidth}	
	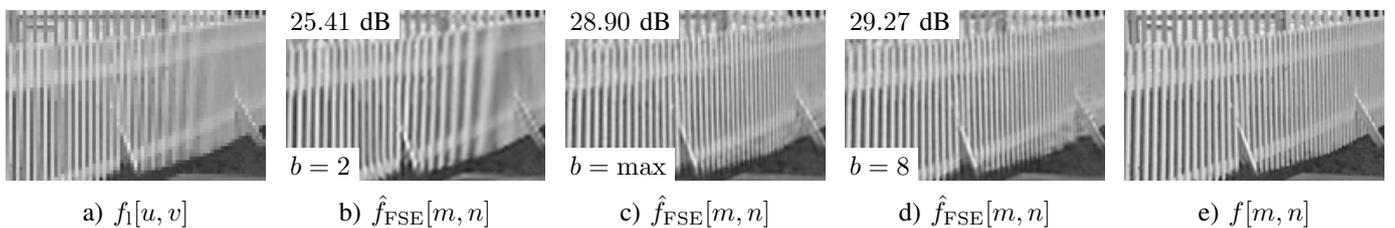
	\vspace*{-0.6cm}
	\caption{Image detail examples of {\it{fence}} reconstructed for non-regular sampling masks with different blocks of size $b$.}
	\label{fig:different_border_width}
	\vspace*{-0.5cm}
\end{figure*}

In Fig.~\ref{fig:image_examples}, image detail examples for three different images ({\it{house, text, roof}}) are illustrated.
The images in Fig.~\ref{fig:image_examples}a represent the images $f_{\mathrm{l}}[u,v]$ that an LR sensor would measure. Fig.~\ref{fig:image_examples}b shows the images $f_{\mathrm{s}}[m,n]$ that a masked LR sensor would acquire followed by HR images $\hat{f}[m,n]$ reconstructed by LIN, SKR, and FSE (Fig.~\ref{fig:image_examples}c,d,e). The original images $f[m,n]$ which are only theoretically available are shown in Fig.~\ref{fig:image_examples}f.
It can be clearly seen that severe aliasing occurs in images that are taken by an LR sensor. Regarding the captured HR images, many pixels have to be reconstructed. 
SKR performs better than LIN on preserving edges. However, an impulsive-like noise impairs the visual quality. Compared to the original images, FSE gives the best visual results.

In Fig.~\ref{fig:different_border_width}, image detail examples of {\it{fence}} reconstructed by FSE for non-regular sampling masks with different blocks of size $b$ are shown. The left one is again the image that an LR sensor would measure and the right one is the original image. In Fig.~\ref{fig:different_border_width}b, the image has been reconstructed for $b=2$. It can clearly be seen that in this case severe aliasing occurs. Using $b=\max$ (Fig.~\ref{fig:different_border_width}c) leads to an almost aliasing free image. Reducing the non-regularity to $b=8$ (Fig.~\ref{fig:different_border_width}d) preserves this visual quality and may give even slightly better results.

The measurements show that it is important to employ an efficient image reconstruction method like FSE for non-regular sampled image data in order to obtain a reasonable HR image quality. Each regarded reconstruction method may benefit from non-regular sampling masks.
Moreover, the manufacturing process may be simplified by using sampling masks that are non-regular only on a smaller scale. These templates can efficiently be stored and then repeated to any sensor size that is needed. Compared to just picking randomly a sampling mask which is non-regular over the whole image sensor size this also leads to a visually noticeable gain in PSNR of up to $0.3$~dB.


\section{Conclusion}
\label{sec:conclusion}
It has been shown that for increasing the spatial resolution by non-regular masking of a low resolution sensor and a subsequent image reconstruction, an efficient reconstruction algorithm like FSE is required in order to obtain a high resolution image. 
Different sampling masks and their influence on image reconstruction have been evaluated.
Reducing the non-regularity by applying $1/4$~sampling only on smaller templates and repeat them to match the sensor size leads to a visually noticeable gain in PSNR compared to arbitrary picked sampling masks that are non-regular over the whole image sensor.
This insight is important for the idea of $1/4$~sampling and future work will cover the design and mathematical description of an ideal sampling mask.

\newpage
\section*{Acknowledgement}
This work has been supported by the Deutsche Forschungsgemeinschaft (DFG) under contract number KA 926/5-1.

\bibliographystyle{IEEEtran}

\bibliography{literature}

\end{document}